\documentclass[useAMS,usenatbib]{mn2e}

\usepackage{amsmath}
\usepackage{amssymb}
\usepackage{graphicx}
\usepackage{dcolumn}
\usepackage{bm}
\usepackage{ulem}

\title[Screening corrections to the Coulomb crystal elastic moduli]
{Screening corrections to the Coulomb crystal elastic moduli}
\author[D. A. Baiko]{D. A. Baiko\thanks{E-mail:baiko@astro.ioffe.ru} \\
A. F. Ioffe Physical-Technical Institute, 
Politekhnicheskaya 26, 194021 St.-Petersburg, Russian Federation}

\begin{document}

\date{Accepted; Received ; in original form}

\pagerange{\pageref{firstpage}--\pageref{lastpage}} \pubyear{2014}

\maketitle

\label{firstpage}

\begin{abstract}
Corrections to elastic moduli, including the effective shear modulus, 
of a solid neutron star crust due to electron screening are calculated.  
At any given mass density the crust is modelled as a body-centered cubic 
Coulomb crystal of fully ionized atomic nuclei of a single type with 
a polarizable charge-compensating electron background. Motion of the 
nuclei is neglected. The electron polarization is described by 
a simple Thomas-Fermi model of exponential electron 
screening. The results of numerical calculations  
are fitted by convenient analytic formulae. 
They should be used for precise neutron star oscillation modelling, 
a rapidly developing branch of stellar seismology. 
\end{abstract}




\section{Introduction}
Discovery of quasi-periodic oscillations (QPO) in soft 
gamma-repeaters \citep{IBS05,SW05,WS06} has stimulated interest in 
studying properties of neutron stars and matter at extreme physical 
conditions by methods of asteroseismology. The QPO are hypothesized 
to be related to neutron star 
vibrations, particularly to torsional vibrations of a solid neutron star 
crust \citep{D98,P05}. The crustal oscillation frequencies are 
determined by elastic moduli of the neutron star crust. 
The main purpose of the present paper is to provide new results
for these quantities under more realistic assumptions about 
the state of the crustal matter than in previous studies.

The bulk of the {\it outer} neutron star crust is made of fully ionized 
ions in crystalline state, immersed in a nearly uniform 
strongly degenerate electron gas. More specifically, we suppose that 
at any given mass 
density all ions have the same charge $Ze$ and mass $M$ and 
form a crystal, if the local temperature $T$ falls below 
the melting temperature $T_{\rm m} = Z^2 e^2 /(a \Gamma_{\rm m})$, 
where $\Gamma_{\rm m} \approx 175$,
and $a=(4 \pi n/3)^{-1/3}$ is the ion sphere radius ($n$ is the ion 
number density, $k_{\rm B}=1$). Typically, one assumes that the ion 
crystal is of the body-centered cubic (bcc) type, as this structure
is preferable thermodynamically for a strictly uniform electron 
background.

In the {\it inner} neutron star crust, at densities above the 
neutron drip density 
$\rho_{\rm d} \approx 4.3 \times 10^{11}$ g cm$^{-3}$, in addition 
to the crystal of ions and electrons, there are neutrons
not bound in the atomic nuclei. The details of the neutron interaction 
with nuclei are not known very well. The motion of nuclei about the
crystal lattice nodes may be affected by the presence 
of neutrons \citep{C12}. However, for the purpose of this paper it is
irrelevant as we will be concentrating on the static lattice case.
At the bottom of the inner crust at densities 
$\rho \ga 10^{14}$ g cm$^{-3}$ there may be a region of nonspherical 
nuclei, known as ``nuclear pasta'', in which the Coulomb crystal 
model fails \citep[e.g.,][]{PP98}. 

The state of the electron subsystem depends on the matter density.  
We shall limit ourselves to such not too low densities, where electrons
are degenerate and ions are completely pressure ionized
\citep[$\rho \ga 10AZ$ g cm$^{-3}$, where $A$ is the number of nucleons 
per nucleus, which is equal to the nucleus mass number in the outer 
crust; see for discussion][]{PR95,HPY07}.
The degenerate electrons are very energetic (at $\rho \gg 10^6$ 
g cm$^{-3}$ they become ultrarelativistic) and it 
is typically considered a good approximation to treat them 
as a constant and uniform charge background. 
In this case, the ion-electron system is called a one-component 
plasma (or a Coulomb crystal, if crystallization is assumed).

However, in reality, 
electrons respond to the ion charge density, which results in a 
screening effect. The strength of this electron polarization
can be characterized by the screening parameter $\kappa a$, 
where $\kappa$ is the Thomas-Fermi wavenumber: 
\begin{equation}
    \kappa a = 2 k_{\rm F} a \sqrt{\alpha/(\pi \beta)}  
    \approx 0.19 Z^{1/3} \beta^{-1/2}~.
\end{equation}
In this case $\beta = v_{\rm F}/c$, $v_{\rm F}$ and $k_{\rm F}$ are 
the electron Fermi velocity and wavevector, respectively,
$\alpha$ is the fine-structure constant, and $c$ is the speed of light. 
Furthermore, $\beta = x_{\rm r}/\sqrt{1+x^2_{\rm r}}$, where 
$x_{\rm r}$ is the electron relativity parameter:
\begin{equation}
     x_{\rm r} = \frac{\hbar k_{\rm F}}{mc} \approx 
     1.009 \left( \frac{\rho_6 Z}{A} \right)^{1/3}~, 
\label{x_def}
\end{equation}
where $m$ is the electron mass and $\rho_6$ is the mass density in units
of $10^6$ g cm$^{-3}$. The electron polarization is not 
very weak even in the inner neutron star crust, where electrons are 
ultrarelativistic. For instance, for $Z=40$,  
$\kappa a \approx 0.65$. In the outer neutron star crust, the
screening parameter decreases with decrease of $Z$ but increases 
with decrease of the mass density (i.e. decrease of the electron 
relativity degree). At sufficiently low density, the 
screening parameter exceeds 1, screening becomes strong and 
full ionization assumption is eventually violated.

The main purpose of this paper is to study the elastic moduli of 
the Coulomb crystal taking into account the electron polarization. 
The groundwork for this problem was laid down 
by \citet{F36}, who calculated the elastic moduli of the static bcc 
Coulomb lattice. \citet{OI90} calculated the elastic moduli of the 
bcc Coulomb crystal taking into account the motion of ions about their 
lattice nodes. They have also introduced in astrophysics the concept 
of shear wave directional averaging and an effective shear modulus
of polycrystalline neutron star crust, which is 
used extensively nowdays. In that work the moduli were found 
numerically
with the aid of Monte Carlo simulations \citep[e.g.,][]{BST66}. 
By the nature of the method, the motion of ions was treated classically.  

\citet{HH08} calculated the effective shear modulus of the Coulomb
crystal taking into account electron screening in the Thomas-Fermi 
model. These authors restricted themselves to 
a specific nucleus charge number and mass density and have not reported
on the dependence of the screening correction on these parameters. 
The calculation was done numerically 
using the molecular dynamics method. Again, the motion of ions was 
strictly classic. 

The effective shear modulus and the Huang elastic shear coefficient 
$S_{1212}$ of the one-component plasma crystal with account
of the ion motion was calculated semi-analytically by \citet{B11}. In 
that paper thermodynamic perturbation theory was employed and 
the ion motion was included in the harmonic lattice model framework. 
Unlike numerical methods of \citet{OI90} and \citet{HH08} 
this approach allowed one to capture quantum effects. The quantum 
effects were found to be rather moderate and of greater importance for  
lighter elements at higher densities. 

The electron polarization correction to the static Coulomb crystal 
effective shear modulus was calculated by \citet{B12}. In this paper, 
screening was described in the linear response formalism. 
Two models of the relativistic electron gas response were considered, 
the Thomas-Fermi model and the zero temperature 
random phase approximation with specific 
formulae for the dielectric function derived by \citet{J62}. 
The results were on average 
compatible with each other, but the calculations with the Jancovici 
screening model revealed sharp singularities in the dependence 
of the effective 
shear modulus on the charge number $Z$ treated 
as a continuous variable. 
The singularities were more pronounced at lower densities. Consequently,  
at certain integer $Z$ and low densities these results deviated 
significantly from the Thomas-Fermi theory. 
At the same time, neither corrections to the individual elastic 
coefficients nor details of the calculations were reported.
Based on the numerical results of \citet{B12}, \citet{KP13} produced 
a fit for the effective shear modulus screening correction in the 
Thomas-Fermi model. 

In their subsequent work, \citet{KP15} have re-evaluated the concept
of the effective shear modulus as it is used in astrophysics. 
Instead of the shear wave directional 
averaging, they have proposed to use a self-consistent theory
\citep[e.g.,][]{dW08}, developed to describe
elastic properties of polycrystalline matter with randomly oriented 
perfect crystallites. In this theory, the effective shear modulus
is given by a nonlinear expression containing all the individual 
elastic moduli of the perfect lattice. 

The goal of the present work is to extend the work of \citet{B12} as 
well as to provide necessary details of these calculations. 
In particular, we report on the screening contributions to all 
second order 
elastic coefficients of the static crystal with the bcc lattice. 
We treat electron polarization perturbatively (perturbative 
calculations 
based on the one-component plasma model fail at 
$\kappa a \gtrsim 1$). We 
propose simple analytic formulae for screening corrections
to all the elastic moduli in the Thomas-Fermi model. We do not 
analyze modification due to screening of the ion motion contribution 
obtained by \citet{B11} as this effect would produce too small 
a correction to the total elastic moduli.    

Besides neutron star crusts, the Coulomb crystals with polarizable 
electron background are expected to form 
in solid cores of white dwarfs, to which the present results also 
apply.

\section{General formalism}
Huang elastic moduli $S_{\alpha \beta \gamma \lambda}$ are the second 
order expansion coefficients of the Helmholtz free energy per unit mass 
in powers of the displacement gradients, 
multiplied by the mass density in the initial, non-deformed 
configuration \citep[e.g., Eq.\ (5.1) of][]{W67}\footnote{This is 
equivalent to differentiating the appropriate thermodynamic potential 
per one ion and multiplying the second derivative
by the non-deformed ion number density.}. They are
also known as the {\it equation of motion} coefficients, as they enter 
the equation of motion of a material, deformed with respect to an 
initial configuration, characterized by an arbitrary uniform stress. 

If a material is under an isotropic initial stress, such as 
hydrostatic pressure, it is more convenient to use Birch elastic 
moduli or the {\it stress-strain} coefficients 
$B_{\alpha \beta \gamma \lambda}$ \citep[in the nomenclature of][]{W67}. 
The convenience stems from the fact that under the {\it isotropic 
initial stress} these coefficients possess the complete Voigt 
symmetry and, at the same time, can be used in the equation of motion 
in place of the Huang coefficients. The Birch coefficients had been  
generalized to the case of a material of an arbitrary symmetry 
under an arbitrary initial stress in \citet{BK65}, where they were 
denoted as $\mathring{c}_{\alpha \beta \gamma \lambda}$. They had been 
reintroduced again in \citet{MMQ02} as second order expansion 
coefficients of the Gibbs free energy in powers of the strain 
parameters. Due to their Voigt symmetry, Voigt notation
\citep[with only two indices, e.g.,][]{W67} is usually used for them. 
For instance, for the cubic symmetry, the only independent coefficients 
are $B_{1111}=\mathring{c}_{1111} = c_{11}$, 
$B_{1122}=\mathring{c}_{1122} = c_{12}$, 
$B_{1212}=\mathring{c}_{1212} = c_{44}$. 
(Let us note, that under an anisotropic initial stress the Birch 
coefficients lose the Voigt symmetry, are no longer equivalent to
the equation of motion coefficients and cannot be obtained from 
the Gibbs free energy, as the latter does not exist.)
  
The relationships between the Huang and Birch coefficients are 
well-known \citep[e.g., Eqs.\ (2.24) and (2.36) of][for an 
arbitrary initial stress or Eqs.\ (2.55) and (2.56) for isotropic 
pressure]{W67}. In particular, for the case of initial isotropic 
pressure $P$ and for a material possessing 
the cubic symmetry $c_{11}=S_{1111}$, $c_{12}=S_{1122}+P$, 
$c_{44}=S_{1212}$. (Additionally, $S_{1221}=S_{1212}+P$.)  
These formulae apply for any partial 
contribution to these coefficients (i.e., due to static lattice, 
electron screening, phonons etc).  

The Huang elastic moduli of static (st) bcc Coulomb 
lattice with a uniform background of opposite charge are \citep[][]{F36}
\begin{eqnarray}
   S^{\rm st}_{1111} &=& - 0.36553930 \, S_0
\nonumber \\
   S^{\rm st}_{1122} &=& - 0.11587344 \, S_0
\nonumber \\
   S^{\rm st}_{1221} &=& - 0.11587344 \, S_0
\nonumber \\
   S^{\rm st}_{1212} &=& S^{\rm st}_{1221} - P^{\rm st} = 
   0.18276965 \, S_0~,
\label{SFuchs}
\end{eqnarray}
where $S_0 = n Z^2 e^2/a$; see also \citet{B11}. The effective 
shear modulus obtained via the directional averaging 
procedure of \citet{OI90} reads
\begin{equation}
  \mu^{\rm st}_{\rm eff} = \frac{1}{5}(S^{\rm st}_{1111}-
  S^{\rm st}_{1122} - S^{\rm st}_{1221}+4S^{\rm st}_{1212}) =
   0.1194572 \, S_0~, 
\label{muFuchs}  
\end{equation}
and the self-consistent shear modulus of \cite{KP15}
becomes (assuming the dominance of the electron bulk modulus
over all the other elastic coefficients)
\begin{equation}
    \mu^{\rm st}_{\rm eff,sc}  
      \approx  0.093 \, S_0~.
\label{muKP}
\end{equation}

Polarizability of the electron background results in a contribution  
$F_\epsilon$ to the system Helmholtz free energy per ion. 
We describe this effect in the linear response formalism 
assuming that the electrons adjust instantaneously to the ion 
configuration. Thus, the electron response can be described
by a static ($\omega=0$) longitudinal dielectric function
$\epsilon(q)$. Assuming further that all ions are fixed at their 
lattice nodes $\bm{R}$ (and thus neglecting modification of the ion 
motion term by the polarization) we obtain:
\begin{eqnarray}
   F_\epsilon &=& \frac{Z^2e^2}{2} 
       \int \frac{{\rm d}\bm{q}}{(2 \pi)^3}
   \frac{4 \pi}{q^2} \left[ \frac{1}{\epsilon(q)} - 1\right]
\nonumber \\ 
     &\times& \left( \sum_{\bm{R}}  \exp{(i \bm{q}\bm{R})} - 
     n \int {\rm d}\bm{r} \exp{(i \bm{q}\bm{r})}
     \right)
\label{Ue}
\end{eqnarray}
\citep[e.g.,][]{HS71,YS89}.  

Introducing uniform deformation with constant 
displacement gradients $u_{\alpha \beta}$,
we replace $R_\alpha \to R_\alpha + u_{\alpha \beta} R_\beta$ and
$r_\alpha \to r_\alpha + u_{\alpha \beta} r_\beta$ in the exponentials
in Eq.\ (\ref{Ue}). Also we have to take into account a dependence
of $\epsilon(q)$ on the density, which changes under the 
deformation by $\delta n \approx n (-u_{\alpha \alpha} 
+ 0.5 u_{\alpha \alpha} u_{\beta \beta} + 0.5 u_{\alpha \beta} 
u_{\beta \alpha})$. [We can also integrate over the new variable 
$r_\alpha + u_{\alpha \beta} r_\beta$ and insert the 
new ion number density
in front of the integral in the second line of Eq.\ (\ref{Ue}), 
but these two modifications cancel each 
other out.] Accordingly, $F_\epsilon \to F_\epsilon 
+ \delta F_\epsilon$, where  $\delta F_\epsilon$ can be expanded
in powers of the displacement gradients: 
\begin{eqnarray}
     n \delta F_\epsilon &=& \frac{n Z^2e^2}{2} 
        \left( \sum_{\bm{R}} - \int n {\rm d} \bm{R} 
        \right)
        \int \frac{{\rm d}\bm{q}}{(2 \pi)^3} \frac{4 \pi}{q^2} 
\nonumber \\
        &\times& \left\{ u_{\alpha \beta}   
       \left( R_\beta \frac{\partial}{\partial R_\alpha}
       - \delta_{\alpha \beta} n  \frac{\partial}{\partial n} \right)
        \right. 
\nonumber \\
     &+& \frac{1}{2} u_{\alpha \beta} u_{\gamma \lambda}   
       \left[ R_\beta R_\lambda  \frac{ \partial^2 }
       {\partial R_\alpha \partial R_\gamma}  
    -  2 \delta_{\gamma \lambda} n R_\beta  
    \frac{\partial^2}{\partial n \partial R_\alpha} \right.   
\nonumber \\
    &+&
       \left.  \left. (\delta_{\alpha \beta} 
       \delta_{\gamma \lambda} + 
       \delta_{\alpha \lambda} \delta_{\beta \gamma}) 
       n \frac{\partial}{\partial n}  
        +  \delta_{\alpha \beta} \delta_{\gamma \lambda}
       n^2  \frac{\partial^2}{\partial n^2}   
       \right]  
        + \ldots \right\} 
\nonumber \\
       &\times& \left[ \frac{1}{\epsilon(q)} - 1 \right]
       \exp{(i \bm{q}\bm{R})}~, 
\label{expans}
\end{eqnarray}
where we have renamed the integration variable as $\bm{R}$ for brevity
of notation, while $\sum_{\bm{R}}$ is over all lattice vectors of the 
non-deformed crystal. The screening (scr) contribution to the isothermal 
Huang elastic coefficient $S^{\rm scr}_{\alpha \beta \gamma \lambda}$  
is the coefficient of the 
$u_{\alpha \beta} u_{\gamma \lambda}/2$ term. The factor 
multiplied by $u_{\alpha \beta}$ 
in the first order term is $-P^{\rm scr} \delta_{\alpha \beta}$, i.e. 
the screening contribution to (minus) pressure. If $[1/\epsilon - 1]$
is replaced by $1$, then the same factors yield the respective 
static lattice contributions. In this case, all the terms containing 
$\partial/\partial n$ vanish. Then it becomes obvious that 
$S^{\rm st}_{1122}=S^{\rm st}_{1221}$ in accordance with 
Eq.\ (\ref{SFuchs}) (this does not hold true for the screening 
contributions). The relationship $S_{1221}=S_{1212}+P$ for both static 
and screening contributions can be seen from the fact
that at fixed $\bm{R}$ the $\bm{q}$-integral (before applying 
the operator in curly brackets) is a function
of $R^2$ only. Then 
\begin{eqnarray}
  R_2^2 \frac{\partial^2}{\partial R_1^2} &=& 4 R_1^2 R_2^2 
   \frac{\partial^2}{\partial (R^2)^2} + 2 R_1^2 
   \frac{\partial}{\partial (R^2)} 
\label{aux1} \\
  R_1 R_2 \frac{\partial^2}{\partial R_1 \partial R_2} &=& 4 R_1^2 R_2^2 
   \frac{\partial^2}{\partial (R^2)^2}
\label{aux2}  \\
   R_1 \frac{\partial}{\partial R_1} &=&  2 R_1^2 
   \frac{\partial}{\partial (R^2)} 
\label{aux3}
\end{eqnarray}
(where subscripts indicate Cartesian components).
In this case, the derivative Eq.\ (\ref{aux1}) contributes to 
$S_{1212}$, the derivative Eq.\ (\ref{aux2}) contributes to 
$S_{1221}$, while the derivative Eq.\ (\ref{aux3}) contributes to 
$-P$. Additionally, $S^{\rm scr}_{1221}$ contains a term with 
$\partial/\partial n$, which is absent in $S^{\rm scr}_{1212}$, but 
is present in $P^{\rm scr}$.   

It is also useful
to write this expansion in reciprocal space:
\begin{eqnarray}
     \delta F_\epsilon &=& \frac{4 \pi n Z^2e^2}{2} 
        \sideset{}{'}\sum_{\bm{G}} 
       \left\{ u_{\alpha \beta}   
       \left( - \frac{\partial}{\partial G_\beta} G_\alpha
       - \delta_{\alpha \beta} n  \frac{\partial}{\partial n} \right)
        \right. 
\nonumber \\
     &+& \frac{1}{2} u_{\alpha \beta} u_{\gamma \lambda}   
       \left[ \frac{ \partial^2 }{\partial G_\beta \partial G_\lambda} 
       G_\alpha G_\gamma  
    +  2 \delta_{\gamma \lambda} n   
    \frac{\partial^2}{\partial n \partial G_\beta} G_\alpha \right.   
\nonumber \\
    &+&
       \left.  \left. (\delta_{\alpha \beta} 
       \delta_{\gamma \lambda} + 
       \delta_{\alpha \lambda} \delta_{\beta \gamma}) 
       n \frac{\partial}{\partial n}  
        +  \delta_{\alpha \beta} \delta_{\gamma \lambda}
       n^2  \frac{\partial^2}{\partial n^2}   
       \right]  
        + \ldots \right\} 
\nonumber \\
        &\times& \frac{1}{G^2} 
        \left[ \frac{1}{\epsilon(G)} - 1 \right] ~. 
\label{Sdef}
\end{eqnarray}
The $\bm{G}$-summation in this formula is over all non-zero 
reciprocal lattice vectors. 

The derivatives featuring in Eq.\ (\ref{Sdef}) can be easily 
evaluated with the following results:
\begin{eqnarray}
  &&  \frac{\partial}{\partial n}
     \left[ \frac{1}{\epsilon(G)} - 1\right] = 
     - \frac{ \dot{\epsilon}}{\epsilon^2}~, 
\nonumber \\
  &&  \frac{\partial^2}{\partial n^2}
     \left[ \frac{1}{\epsilon(G)} - 1\right] = 
     \frac{ 2 \dot{\epsilon}^2}{\epsilon^3} -
     \frac{ \ddot{\epsilon}}{\epsilon^2}~, 
\nonumber \\
  && \frac{\partial}{\partial G_\beta}
     \frac{G_\alpha}{G^2} 
     \left[ \frac{1}{\epsilon(G)} - 1\right]
\nonumber \\     
   && ~~ = ~
     \left( \frac{\delta_{\alpha \beta}}{G^2} - 
     \frac{2 G_\alpha G_\beta}{G^4} \right)
     \left[ \frac{1}{\epsilon(G)} - 1\right] -
     \frac{G_\alpha G_\beta}{G^3} \frac{\epsilon'}{\epsilon^2}~,
\nonumber \\
  && \frac{\partial^2}{\partial G_\beta \partial G_\lambda}
     \frac{G_\alpha G_\gamma}{G^2} 
     \left[ \frac{1}{\epsilon(G)} - 1\right] 
\nonumber \\     
   && ~~ = ~
     (\delta_{\alpha \lambda} \delta_{\beta \gamma} + 
     \delta_{\alpha \beta} \delta_{\gamma \lambda}) 
     \frac{1}{G^2}\left(\frac{1}{\epsilon} - 1 \right)   
\nonumber \\     
     && ~~ - ~
            (\delta_{\beta \lambda} G_\alpha G_\gamma +
                   \delta_{\beta \gamma} G_\alpha G_\lambda +
                   \delta_{\alpha \beta} G_\gamma G_\lambda 
\nonumber \\                   
     && ~~ + ~ 
                   \delta_{\alpha \lambda} G_\beta G_\gamma +
                   \delta_{\gamma \lambda} G_\alpha G_\beta ) 
      \frac{1}{G^2}
      \left[\frac{2}{G^2} \left(\frac{1}{\epsilon} - 1 \right) + 
      \frac{\epsilon'}{G \epsilon^2} \right]       
\nonumber \\
     && ~~ + ~ \frac{G_\alpha G_\beta G_\gamma G_\lambda}{G^4}
      \left[\frac{8}{G^2} \left(\frac{1}{\epsilon} - 1 \right)
      +\frac{5 \epsilon'}{G \epsilon^2} + \frac{2 \epsilon'^2}{\epsilon^3}
      -\frac{\epsilon''}{\epsilon^2} \right],
\nonumber \\
  && \frac{\partial^2}{\partial n \partial G_\beta}
     \frac{G_\alpha}{G^2} 
     \left[ \frac{1}{\epsilon(G)} - 1\right]
\nonumber \\
   && ~~ = ~
       - \frac{\delta_{\alpha \beta} 
       \dot{\epsilon}}{G^2 \epsilon^2} + \frac{G_\alpha G_\beta}{G^2}
       \left(\frac{2 \dot{\epsilon}}{G^2 \epsilon^2}
       + \frac{2 \dot{\epsilon} \epsilon'}{G \epsilon^3} 
       - \frac{\dot{\epsilon}'}{G \epsilon^2} \right), 
\label{Sres}
\end{eqnarray}
where a dot over $\epsilon$ means the derivative with respect to $n$ 
(the electron density is $Zn$), while a prime over $\epsilon$ means 
the derivative with respect to $G$.

\section{Calculations with the Thomas-Fermi dielectric function}
The Thomas-Fermi (TF) dielectric function
$\epsilon_{\rm TF}(q)=1+\kappa^2/q^2$ describes 
exponential screening of the Coulomb potential with the screening 
length equal to $1/\kappa$. 
In principle, one can use Eqs.\ (\ref{Sdef}) and (\ref{Sres}) to find 
the screening contributions to the Huang coefficients. However,
$(1/\epsilon_{\rm TF}) -1 \sim q^{-2}$ at large $q$ and the expression 
under the three-dimensional reciprocal lattice sum would decay 
only as $q^{-4}$. Consequently, the convergence would be very slow.   

Fortunately, the problem can be reformulated using the ``screened'' 
Ewald technique identical to that used in \cite{B02}. 
The final practical formula  
for $S^{\rm scr}_{\alpha \beta \gamma \lambda}$
in the Thomas-Fermi approximation 
is given in the Appendix, Eq.\ (\ref{Spract}). Its derivation
is based on several simple ideas. 
First, we note that a 
contribution to $S^{\rm scr}_{\alpha \beta \gamma \lambda}$, 
corresponding to a given $\bm{R}$, can be written as
\begin{eqnarray}
     && \frac{Z^2e^2}{2} \left[ R_\beta R_\lambda 
        \frac{\partial^2}{\partial R_\alpha \partial R_\gamma}
        - 2 \delta_{\gamma \lambda} 
        n R_\beta 
        \frac{\partial^2}{\partial n \partial R_\alpha} 
         \right.
\nonumber \\
        &+& \left. (\delta_{\alpha \beta} 
       \delta_{\gamma \lambda} + 
       \delta_{\alpha \lambda} \delta_{\beta \gamma}) 
       n \frac{\partial}{\partial n}  
        +  \delta_{\alpha \beta} \delta_{\gamma \lambda}
       n^2  \frac{\partial^2 }{\partial n^2} 
       \right]
\nonumber \\
       &\times& \frac{1}{R} [\exp{(-\kappa R)} - 1].
\label{Rdiff}
\end{eqnarray}
[Interestingly, the $\bm{R} = 0$ term produces non-zero 
contributions to
$S^{\rm scr}_{1111}$, $S^{\rm scr}_{1122}$ and $S^{\rm scr}_{1221}$,
but not to the shear coefficient $S^{\rm scr}_{1212}$.
This is the $S^{(1)}$ term in Eq.\ (\ref{Spract}).]  

Then we use the following standard integral:
\begin{equation}
 \frac{e^{-\kappa R} - 1}{R} = \frac{2}{\sqrt{\pi}} 
  \int^\infty_0 {\rm d} \rho e^{-\rho^2 R^2} 
  \left(e^{- \kappa^2/(4 \rho^2)} - 1\right)~,
\label{standard}
\end{equation}
and split it as 
$\int^\infty_0 = \int^{\cal A}_0 + \int^\infty_{\cal A}$, 
where $0<{\cal A}<\infty$ 
but otherwise arbitrary. The integral 
$\int^\infty_{\cal A}$ may be expressed via the 
complementary error
functions ${\rm erfc}({\cal A}R \pm \kappa/(2{\cal A}))$ 
which decay 
very rapidly at large $R$. We differentiate them
as prescribed by Eq.\ (\ref{Rdiff}) and this results in contributions
$S^{(2)}$ and $S^{(3)}$ in Eq.\ (\ref{Spract}) 
for $\int{\rm d}\bm{R}$ and 
$\sum_{\bm{R}\ne 0}$, respectively. In $S^{(3)}$ it is sufficient to 
sum over very few first shells of lattice vectors $\bm{R}$ 
to achieve convergence.  

\begin{figure}
\begin{center}
\leavevmode
\includegraphics[bb=3 23 346 348, width=84mm]{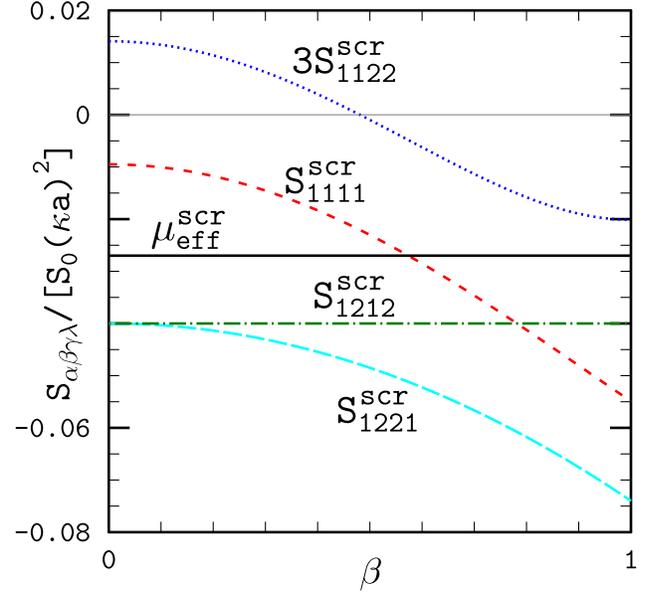}
\end{center}
\vspace{-0.4cm}
\caption[ ]{Screening corrections to elastic moduli calculated 
using $\epsilon_{\rm TF}$.}
\label{stf}
\end{figure}

Clearly, this procedure relies on the fact that 
${\cal A}>0$, and a different treatment of the integral 
$\int^{\cal A}_0$ is needed. 
We substitute the well-known formula
\begin{equation}
       \exp{\left(-\rho^2 R^2\right)} 
       = \int \frac{{\rm d}\bm{q}}{(2 \pi)^3} 
       \frac{\pi \sqrt{\pi}}{\rho^3} 
       \exp{\left[i \bm{q}\bm{R}-q^2/(4\rho^2)\right]} 
\end{equation}
in Eq.\ (\ref{standard}). After that, for instance, 
$R_\beta R_\lambda \partial^2/(\partial R_\alpha \partial R_\gamma)$ 
in Eq.\ (\ref{Rdiff}) becomes
$q_\alpha q_\gamma \partial^2/(\partial q_\beta \partial q_\lambda)$.
Summation over $\bm{R}$ then yields a series of delta-functions 
in reciprocal space via the identity
$\sum_{\bm{R}} \exp{(i\bm{q}\bm{R})} = 
(2\pi)^3 n \sum_{\bm{G}} \delta(\bm{q}-\bm{G})$. 
[The $\bm{R}=0$ term now has to be subtracted as it is already present
in the form of $S^{(1)}$. This results in the $S^{(4)}$ term in 
Eq.\ (\ref{Spract}). If ${\cal A} \to \infty$, $S^{(4)} \to - S^{(1)}$ 
as it should be.] The $\bm{q}$-integral 
is then taken with the aid of the integration by parts and the 
delta-functions. The remaining $\int^{\cal A}_0 {\rm d} \rho$ 
turns out to
be elementary and the result (the $S^{(5)}$ term) contains the 
rapidly decaying 
(for ${\cal A}<\infty$) function 
$\exp{[-(G^2+\kappa^2)/(4 {\cal A}^2)]}$. It is again 
sufficient to sum over 
very few first shells of reciprocal lattice vectors $\bm{G}$ 
to achieve convergence.

Using Eq.\ (\ref{Spract}), we have calculated the screening corrections
to all Huang elastic coefficients.    
Since these are perturbative calculations, only 
the lowest order terms in $\kappa a$ are described correctly. They
are $\propto (\kappa a)^2$ and our main results 
can be summarized as
\begin{eqnarray}
   S^{\rm scr}_{1111} &=& -( 0.0095 + 0.057 \beta^2 - 0.0116 \beta^4) 
   \, (\kappa a)^2 \, S_0~,  
\nonumber \\
   S^{\rm scr}_{1122} &=& (0.0047 -0.023 \beta^2+0.0116 \beta^4) 
   \, (\kappa a)^2 \, S_0~, 
\nonumber \\
   S^{\rm scr}_{1221} &=&  -(0.04 + 0.034 \beta^2) \, (\kappa a)^2 \, 
    S_0~, 
\nonumber \\
   S^{\rm scr}_{1212} &=& c^{\rm scr}_{44} = - 0.04 \, (\kappa a)^2 \, 
   S_0~.
\label{Sscr_num}
\end{eqnarray}
The dependence of some coefficients on $\beta$ stems from 
the dielectric function dependence on $n$ and variation of the latter
for certain types of deformations. These coefficients
are shown in Fig.\ \ref{stf}. The $S^{\rm scr}_{1122}$ coefficient
is additionally multiplied by $3$ to improve the figure readability.
The screening correction to the effective shear modulus 
obtained via the averaging procedure is
\begin{eqnarray}
      \mu^{\rm scr}_{\rm eff} &=& 
      \frac{1}{5}(S^{\rm scr}_{1111}-S^{\rm scr}_{1122}
      -S^{\rm scr}_{1221}+4S^{\rm scr}_{1212})
\nonumber \\
      &=& - 0.027 \, (\kappa a)^2 \, S_0~.
\label{mufit}
\end{eqnarray}
Finally, we have expanded the nonlinear expression of \citet{KP15}
for the self-consistent effective shear modulus and obtained the 
screening correction to it as 
\begin{equation}
     \mu^{\rm scr}_{\rm eff,sc} = -0.022 \, (\kappa a)^2 \, S_0~.
\label{muscfit}
\end{equation}
The polarization contributions to the Huang shear coefficient
$S_{1212}$ and the effective shear moduli are, 
as expected, negative. This means that screening reduces lattice
resistance to a shear strain. Also they do not contain contributions
associated with the density dependence of the dielectric function. 
Naturally, the polarization corrections
Eqs.\ (\ref{Sscr_num})---(\ref{muscfit}) are 
smaller than the
electrostatic elastic moduli Eqs.\ (\ref{SFuchs})---(\ref{muKP}) 
within the limits
of validity of the perturbative treatment of screening. 
The fit (\ref{mufit}) has already been proposed 
by \citet{KP13} based on the numerical data of \citet{B12}. However,
their coefficient in Eq.\ (\ref{mufit}) is $\sim 20\%$ greater 
than ours due to a minor numerical error on their part. 
Results (\ref{Sscr_num}) and (\ref{muscfit}) are new.

It is also convenient to reformulate our results in terms of 
the pressure 
\begin{equation}
    P^{\rm scr} = S^{\rm scr}_{1221} - S^{\rm scr}_{1212} = 
    - 0.034 \beta^2 \, (\kappa a)^2 \, S_0~,
\end{equation}
bulk modulus
\begin{eqnarray}
     \frac{1}{3} (c^{\rm scr}_{11}+ 2 c^{\rm scr}_{12}) &=&
     \frac{1}{3} (S^{\rm scr}_{1111} + 2 S^{\rm scr}_{1122} 
     + 2 S^{\rm scr}_{1221} - 2 S^{\rm scr}_{1212})
\nonumber \\
  &=&   -(0.057 \beta^2 - 0.0116 \beta^4) \, (\kappa a)^2 \, S_0~,  
\end{eqnarray}
and
\begin{eqnarray}
  c^{\rm scr}_{11} - c^{\rm scr}_{12} &=&
  S^{\rm scr}_{1111} - S^{\rm scr}_{1122} - S^{\rm scr}_{1221} 
  + S^{\rm scr}_{1212} 
\nonumber \\  
  &=& - 0.0142 \, (\kappa a)^2 \, S_0~.
\end{eqnarray}

From the results reported so far one may gain an impression that 
the screening corrections can become very large relative to the pure 
Coulomb values of the elastic moduli at low mass densities 
and at $Z \gg 1$. In this regime our perturbative method 
does not work. For instance, at 
$x_{\rm r}=0.2$ and $Z=26$, $\kappa a \approx 1.3$.  
However, in this regime we expect an onset of partial ionization. In 
this case, matter can be approximately described as a Coulomb system
with an effective ion charge $Z_{\rm eff}<Z$. This effective charge will
result in reduced effective values of $\kappa a$ and $S_0$ to
be used in Eqs.\ (\ref{SFuchs})---(\ref{muKP}) 
and (\ref{Sscr_num})---(\ref{muscfit}).

\section{Conclusion}
We have calculated electron polarization corrections to elastic moduli
of Coulomb crystals in neutron star crust. The effect was described
by the Thomas-Fermi model of exponential electron screening.

Combining with Eqs.\ (45) and (46) of \citet{B11}, 
we can now write expressions for the total effective shear moduli 
$\mu^{\rm tot}_{\rm eff}$, $\mu^{\rm tot}_{\rm eff,sc}$ 
and the elastic shear coefficient $S^{\rm tot}_{1212}$ including the 
effects of ion motion and electron screening (unfortuantely, 
the ion motion contribution is presently not available for 
$\mu^{\rm tot}_{\rm eff,sc}$):
\begin{eqnarray}
   S^{\rm tot}_{1212} &=& 
  [0.18276965 - 0.04 (\kappa a)^2] \, S_0 
\nonumber \\
  &-& 
  \left[0.2057 + 439 \left( \frac{T}{T_{\rm p}}\right)^3\right]^{1/3} 
  n T_{\rm p}~,
\nonumber \\
  \mu^{\rm tot}_{\rm eff} &=& 
  [0.1194572 - 0.027 (\kappa a)^2] \, S_0 
\nonumber \\
  &-& 
  \left[0.05008 +136.6 \left( \frac{T}{T_{\rm p}}\right)^3\right]^{1/3} 
  n T_{\rm p}~,
\nonumber \\
    \mu^{\rm tot}_{\rm eff,sc} &=& [0.093 - 0.022 (\kappa a)^2] \, S_0~, 
\label{totals}
\end{eqnarray}
where $T_{\rm p} = \hbar \sqrt{4 \pi n Z^2 e^2 / M} \approx
7.8 \times 10^6 (Z/A) \sqrt{\rho_6 X}$~K is the ion plasma temperature.
In this case $X=A/A_{\rm N}$, where $A_{\rm N}$ is the number of 
nucleons {\it bound} in a nucleus ($A_{\rm N}=A$ in the outer crust).
We would like to emphasize that Eqs.\ (\ref{totals}) do not take into 
account
details of ``free'' neutron interactions with nuclei in the inner 
neutron star crust. Here it is assumed that these effects result in 
an effective (increased) nucleus mass and a renormalized plasma 
frequency.

\section*{Acknowledgments}
The author is really thankful to the anonymous referee for numerous 
comments, which led to a significant improvement of the manuscript. 
The author is also grateful to C.J.\ Pethick and D.G.\ Yakovlev for 
discussions. This work was supported by RSF, grant No.\ 14-12-00316.

\appendix

\section{}
In the Thomas-Fermi approximation
\begin{eqnarray}
   S^{\rm scr}_{\alpha \beta \gamma \lambda} &=&
   \frac{nZ^2e^2}{2} \sum_{i=1}^5 S^{(i)}_{\alpha \beta \gamma \lambda}~,
\label{Spract} \\
     S^{(1)}_{\alpha \beta \gamma \lambda} &=&  
     - \delta_{\alpha \beta} \delta_{\gamma \lambda} 
     (\kappa^1_1+\kappa^1_2) 
     - \delta_{\alpha \lambda} \delta_{\beta \gamma} \kappa^1_1~,
\nonumber \\
     S^{(2)}_{\alpha \beta \gamma \lambda} &=&
     -\frac{\pi n}{{\cal A}^2} 
     \left[\delta_{\alpha \beta} \delta_{\gamma \lambda} 
     \left(1+3n\frac{{\rm d}}{{\rm d} n} + 
     n^2\frac{{\rm d}^2}{{\rm d} n^2} \right) \right.
\nonumber \\
     &+& \left. \delta_{\alpha \lambda} \delta_{\beta \gamma}
     \left(1+n\frac{{\rm d}}{{\rm d} n}\right) \right] F(u)~,
\nonumber \\
  S^{(3)}_{\alpha \beta \gamma \lambda} &=& 
  \sideset{}{'}\sum_{\bm{R}}\left[R_\beta R_\lambda 
  \frac{\partial^2}{\partial R_\alpha \partial R_\gamma}
  - 2 \delta_{\gamma \lambda} n R_\beta 
  \frac{\partial^2}{\partial n \partial R_\alpha} \right.
\nonumber \\
  &+& \left. (\delta_{\alpha \beta} \delta_{\gamma \lambda} +
  \delta_{\alpha \lambda } \delta_{\beta \gamma}) 
  n \frac{\partial}{\partial n} +
  \delta_{\alpha \beta} \delta_{\gamma \lambda} 
  n^2 \frac{\partial^2}{\partial n^2} \right]
\nonumber \\
  &\times& \left\{ \frac{{\cal A}}{2x} [E_+(x,y)+E_-(x,y)] 
  - \frac{{\cal A}}{x} {\rm erfc}(x)
  \right\}~,
\nonumber \\
  S^{(4)}_{\alpha \beta \gamma \lambda} &=&
  \frac{\delta_{\alpha \beta} \delta_{\gamma \lambda}}{2 \kappa} 
  \left[ E_+(0,y) 
  \left(\kappa^2_1 + \kappa^2_2 - 
  \frac{\kappa^2_1 \kappa^1_1}{\kappa}   \right) \right. 
\nonumber \\
  &-& \left. \frac{\kappa^2_1 \kappa^1_1}{{\cal A} \sqrt{\pi}}
  \exp{(-u)} \right] +
  \frac{\delta_{\alpha \lambda} \delta_{\beta \gamma}}{2 \kappa} E_+(0,y) 
  \kappa^2_1~,  
\nonumber \\
  S^{(5)}_{\alpha \beta \gamma \lambda} &=&
  \frac{\pi n}{{\cal A}^2} \sideset{}{'}\sum_{\bm{G}}
  \left[ \frac{\partial^2}{\partial G_\beta \partial G_\lambda} 
  G_\alpha G_\gamma + 2 \delta_{\gamma \lambda} n G_\alpha  
  \frac{\partial^2}{\partial n \partial G_\beta}  \right. 
\nonumber \\
  &+& \left. (3 \delta_{\alpha \beta} \delta_{\gamma \lambda} +
  \delta_{\alpha \lambda } \delta_{\beta \gamma}) 
  n \frac{\partial}{\partial n} +
  \delta_{\alpha \beta} \delta_{\gamma \lambda} 
  n^2 \frac{\partial^2}{\partial n^2} 
  \right] 
\nonumber \\
  &\times& [D(g,u)-D(g,0)]~,
\nonumber 
\end{eqnarray}
where primes mean that the $\bm{R}=0$ and $\bm{G}=0$ terms in the 
lattice sums must be omitted. Furthermore,
\begin{eqnarray}
       F(u) &=& \frac{1}{u}\left[1- \exp{(-u)}\right] - 1~,
\nonumber \\
     E_{\pm}(x,y) &=& \exp{(\pm 2 x y )} 
     {\rm erfc}(x \pm y)~, 
\nonumber \\
     {\rm erfc}(x) &=& E_\pm(x,0)~,
\nonumber \\
      D(g,u) &=& \frac{1}{g+u} \exp{(-g-u)}~,
\nonumber \\
      u&=&y^2~, ~~~ x={\cal A} R~, 
\nonumber \\
       y&=&\frac{\kappa}{2{\cal A}}~, ~~~ g=\frac{G^2}{4{\cal A}^2}~, 
\nonumber
\end{eqnarray}
\begin{eqnarray}
      n\frac{{\rm d} F}{{\rm d} n}  &=& 
      \frac{{\rm d}F}{{\rm d}u} \frac{\kappa^2_1}{4{\cal A}^2}~,
\nonumber \\
      n\frac{{\rm d}^2 F}{{\rm d} n^2}  &=&
      \frac{{\rm d}^2 F}{{\rm d}u^2} 
      \frac{(\kappa^2_1)^2}{16 {\cal A}^4}
      + \frac{{\rm d} F}{{\rm d}u} \frac{\kappa^2_2}{4 {\cal A}^2}~,
\nonumber
\end{eqnarray}
\begin{eqnarray}
    \frac{\partial^2}{\partial R_\alpha \partial R_\gamma} 
   \frac{E_\pm}{x} ~~=~~ \delta_{\alpha \gamma} 
   \frac{{\cal A}^2}{x^2} \left(\frac{\partial E_\pm}{\partial x} -
   \frac{E_\pm}{x} \right) && 
\nonumber \\
 + R_\alpha R_\gamma  
   \frac{{\cal A}^4}{x^3} \left(\frac{\partial^2 E_\pm}{\partial x^2} -
   \frac{3}{x} \frac{\partial E_\pm}{\partial x}  
   +  \frac{3}{x^2} E_\pm \right)~, &&
\nonumber
\end{eqnarray}
\begin{eqnarray}
    n \frac{\partial^2}{\partial n \partial R_\alpha} 
   \frac{E_\pm}{x} &=& \frac{{\cal A} R_\alpha \kappa^1_1}{2 x^2}
   \left(\frac{\partial^2 E_\pm}{\partial x \partial y} 
   - \frac{1}{x} \frac{\partial E_\pm}{\partial y} \right)~, 
\nonumber \\
     n \frac{\partial}{\partial n} \frac{E_\pm}{x} &=&
    \frac{\partial E_\pm}{\partial y} \frac{\kappa_1^1}{2 {\cal A} x}~,
\nonumber \\
     n^2 \frac{\partial^2}{\partial n^2} \frac{E_\pm}{x} &=&
    \frac{\partial^2 E_\pm}{\partial y^2} \frac{(\kappa_1^1)^2}
    {4 {\cal A}^2 x} + \frac{\partial E_\pm}{\partial y}
    \frac{\kappa_2^1}{2 {\cal A} x}~,
\nonumber
\end{eqnarray}
with all $n$-derivatives of $E_\pm(x,0)$ being 0,
\begin{eqnarray}
     \frac{\partial E_\pm}{\partial x} &=& \pm 2 y E_\pm - E_R~,
\nonumber \\
     \frac{\partial E_\pm}{\partial y} &=& \pm 2 x E_\pm \mp E_R,
\nonumber \\
     \frac{\partial^2 E_\pm}{\partial x^2} &=& 4 y^2 E_\pm \mp 2 y E_R
    + 2 x E_R~,
\nonumber \\
     \frac{\partial^2 E_\pm}{\partial y^2} &=& 4 x^2 E_\pm - 2 x E_R
    \pm 2 y E_R~,
\nonumber \\
     \frac{\partial^2 E_\pm}{\partial x \partial y} &=& 
    E_\pm (4 x y \pm 2)~,
\nonumber \\
     E_R &=& \frac{2}{\sqrt{\pi}} \exp{(-x^2-y^2)}~,
\nonumber
\end{eqnarray}
\begin{eqnarray}
  \frac{\partial^2 (G_\alpha G_\gamma D)}
  {\partial G_\beta \partial G_\lambda} 
   = (\delta_{\alpha \beta} \delta_{\gamma \lambda} + 
  \delta_{\alpha \lambda } \delta_{\beta \gamma}) D +
  G_\alpha G_\gamma \frac{\partial^2 D}
  {\partial G_\beta \partial G_\lambda} && 
\nonumber \\ 
 +  \delta_{\alpha \beta} G_\gamma \frac{\partial D}{\partial G_\lambda}   
 +  \delta_{\alpha \lambda} G_\gamma \frac{\partial D}{\partial G_\beta} 
 +  \delta_{\gamma \beta} G_\alpha \frac{\partial D}{\partial G_\lambda}   
 +  \delta_{\gamma \lambda} G_\alpha \frac{\partial D}{\partial G_\beta}~,   
 &&  
\nonumber
\end{eqnarray}
\begin{eqnarray}
      \frac{\partial D}{\partial G_\beta} &=&
      \frac{G_\beta}{2 {\cal A}^2}
       \frac{\partial D}{\partial g}~, 
\nonumber \\
       \frac{\partial^2 D}{\partial G_\beta \partial G_\lambda}  &=&
       \frac{\delta_{\beta \lambda}}{2 {\cal A}^2} 
       \frac{\partial D}{\partial g}  
       + \frac{G_\beta G_\lambda}{4 {\cal A}^4} 
       \frac{\partial^2 D}{\partial g^2}~,
\nonumber \\
    n   \frac{\partial^2 D}{\partial n \partial G_\beta}  &=&
       \frac{G_\beta \kappa^2_1}{8 {\cal A}^4} 
       \frac{\partial^2 D}{\partial g^2}~, 
\nonumber \\
      n   \frac{\partial D}{\partial n}  &=&        
      \frac{\kappa^2_1}{4 {\cal A}^2} \frac{\partial D}{\partial g}~, 
\nonumber \\
      n^2   \frac{\partial D}{\partial n^2}  &=&        
      \frac{\kappa^2_2}{4 {\cal A}^2} \frac{\partial D}{\partial g} 
   + \frac{(\kappa^2_1)^2}{16 {\cal A}^4} 
   \frac{\partial^2 D}{\partial g^2}~, 
\nonumber 
\end{eqnarray}
with all $n$-derivatives of $D(g,0)$ being 0,
\begin{eqnarray}
       \kappa^1_1 &=& n\frac{{\rm d} \kappa}{{\rm d} n} = 
       \frac{\kappa (1+\beta^2)}{6}~,           
\nonumber \\
       \kappa^1_2 &=& n^2 \frac{{\rm d}^2 \kappa}{{\rm d} n^2} = 
       - \frac{\kappa (5+3\beta^4)}{36}~,           
\nonumber \\     
       \kappa^2_1 &=& n\frac{{\rm d} \kappa^2}{{\rm d} n} = 
       \frac{\kappa^2 (1+\beta^2)}{3}~,           
\nonumber \\
       \kappa^2_2 &=& n^2 \frac{{\rm d}^2 \kappa^2}{{\rm d} n^2} = 
       - \frac{\kappa^2 (2-\beta^2+\beta^4)}{9}~.           
\nonumber
\end{eqnarray}

\label{lastpage}


\begin{thebibliography}{}

\bibitem[\protect\citeauthoryear{Baiko}{2002}]{B02}
      Baiko D.A., 2002, Phys. Rev. E, 66, 056405

\bibitem[\protect\citeauthoryear{Baiko}{2011}]{B11} 
      Baiko D.A., 2011, MNRAS, 416, 22

\bibitem[\protect\citeauthoryear{Baiko}{2012}]{B12} 
      Baiko D.A., 2012, Contrib. Plasma Phys., 52, 157

\bibitem[\protect\citeauthoryear{Barron \& Klein}{1965}]{BK65}
      Barron T.H.K., Klein M.L., 1965,
      Proc. Phys. Soc., 85, 523

\bibitem[\protect\citeauthoryear{Brush, Sahlin \& Teller}{Brush et al.}{1966}]{BST66}
      Brush S.G., Sahlin H.L., Teller E., 1966,
      J. Chem. Phys., 45, 2102

\bibitem[\protect\citeauthoryear{Chamel}{2012}]{C12}
      Chamel N., 2012, Phys. Rev. C, 85, 035801

\bibitem[\protect\citeauthoryear{deWit}{2008}]{dW08}
      deWit R., 2008, J. Mech. Mater. Struct., 3, 195

\bibitem[\protect\citeauthoryear{Duncan}{1998}]{D98}
      Duncan R.C., 1998, ApJ, 498, L45

\bibitem[\protect\citeauthoryear{Fuchs}{1936}]{F36}
      Fuchs K., 1936, Proc. Roy. Soc. London, 153, 622

\bibitem[\protect\citeauthoryear{Haensel, Potekhin \& Yakovlev}{2007}]{HPY07}
      Haensel P., Potekhin A.Y., Yakovlev D.G., 2007, 
      Neutron Stars 1: Equation of State and Structure. Springer, New York 

\bibitem[\protect\citeauthoryear{Horowitz \& Hughto}{2008}]{HH08}
      Horowitz C.J., Hughto J., 2008, preprint (astro-ph/0812.2650)

\bibitem[\protect\citeauthoryear{Hubbard \& Slattery}{1971}]{HS71}
      Hubbard W.B., Slattery W.L., ApJ, 168, 131

\bibitem[\protect\citeauthoryear{Israel et al.}{2005}]{IBS05}
      Israel G.L. et al., 2005, ApJ, 628, L53

\bibitem[\protect\citeauthoryear{Jancovici}{1962}]{J62}
      Jancovici B., 1962, Nuov. Cim., 25, 428

\bibitem[\protect\citeauthoryear{Kobyakov \& Pethick}{2013}]{KP13}
      Kobyakov D., Pethick C.J., 2013, Phys. Rev. C, 87, 055803 
 
\bibitem[\protect\citeauthoryear{Kobyakov \& Pethick}{2015}]{KP15}
      Kobyakov D., Pethick C.J., 2015, MNRAS, 449, L110 
 
\bibitem[\protect\citeauthoryear{Marcus, Hong Ma \& Qiu}{Marcus et al.}{2002}]{MMQ02}
      Marcus P.M., Hong Ma, Qiu S.L., 2002,
      J. Phys.: Cond. Mat., 14, L525

\bibitem[\protect\citeauthoryear{Ogata \& Ichimaru}{1990}]{OI90}
      Ogata S., Ichimaru S., 1990, Phys. Rev. A, 42, 4867

\bibitem[\protect\citeauthoryear{Pethick \& Potekhin}{1998}]{PP98}
      Pethick C.J., Potekhin A.Y., 1998, Phys. Lett. B, 427, 7

\bibitem[\protect\citeauthoryear{Pethick \& Ravenhall}{1995}]{PR95}
      Pethick C.J., Ravenhall D.G., 1995, Annu. Rev. Nucl. Part. Sci., 
      45, 429

\bibitem[\protect\citeauthoryear{Piro}{2005}]{P05}
      Piro A.L., 2005, ApJ, 634, L153
   
\bibitem[\protect\citeauthoryear{Strohmayer \& Watts}{2005}]{SW05}
     Strohmayer T.E., Watts A.L., 2005, ApJ, 632, L111

\bibitem[\protect\citeauthoryear{Wallace}{1967}]{W67}
     Wallace D.C., 1967, Phys. Rev., 162, 776

\bibitem[\protect\citeauthoryear{Watts \& Strohmayer}{2006}]{WS06}
     Watts A.L., Strohmayer T.E., 2006, ApJ, 637, L117

\bibitem[\protect\citeauthoryear{Yakovlev \& Shalybkov}{1989}]{YS89}
      Yakovlev D.G., Shalybkov D.A., 1989, Sov. Sci. Rev., 
      7, 311 

\end{thebibliography}
\end{document}